\documentclass{elsart}

 \usepackage{graphicx}

\usepackage{amssymb}

\begin{document}

\begin{frontmatter}



\title{Results on the Coherent Interaction of High Energy Electrons and 
Photons in Oriented Single Crystals}


\author[northwestern]{A.~Apyan\corauthref{cor}},
\corauth[cor]{Corresponding author. Tel: 1 847-467-5965; Fax: 
1 847-467-6857}
\ead{aapyan@lotus.phys.northwestern.edu}
\author[yerevan]{R.O.~Avakian},
\author[uppsala]{B.~Badelek},
\author[firenze]{S.~Ballestrero},
\author[torino]{C.~Biino},
\author[northwestern]{I.~Birol},
\author[perugia]{P.~Cenci},
\author[scholand]{S.H.~Connell},
\author[northwestern]{S.~Eichblatt},
\author[northwestern]{T.~Fonseca},
\author[grenoble]{A.~Freund},
\author[cern]{B.~Gorini},
\author[scholand]{R.~Groess},
\author[yerevan]{K.~Ispirian},
\author[nikhef]{T.J.~Ketel},
\author[kurchatov]{Yu.V.~Kononets},
\author[compostela]{A.~Lopez},
\author[firenze]{A.~Mangiarotti},
\author[nikhef]{B.~van~Rens},
\author[scholand]{J.P.F.~Sellschop\thanksref{dec}},
\thanks[dec]{Deceased}
\author[northwestern]{M.~Shieh},
\author[firenze]{P.~Sona},
\author[budker]{V.~Strakhovenko},
\author[isr]{E.~Uggerh{\o}j},
\author[aarhus]{U.I.~Uggerh{\o}j},
\author[california]{G.~Unel},
\author[northwestern]{M.~Velasco},
\author[capetown]{Z.Z.~Vilakazi}
\author[uppsala]{and O.~Wessely}

\address[northwestern]{Northwestern University, Dept. of Physics and Astronomy, 
2145 Sheridan Road, Evanston, Illinois, USA}
\address[yerevan]{Yerevan Physics Institute, Yerevan 375036, Armenia}
\address[uppsala]{Uppsala University, Uppsala 75105, Sweden}
\address[firenze]{INFN and University of Firenze, Firenze 50121, Italy}
\address[torino]{INFN and University of Torino, Torino 10129, Italy}
\address[perugia]{INFN, Perugia 06123, Italy}
\address[scholand]{Schonland Research Institute - University of the
Witwatersrand, Johannesburg 2050, South Africa}
\address[grenoble]{ESRF, Grenoble 38043, France}
\address[cern]{CERN, Geneva 1211, Switzerland}
\address[nikhef]{NIKHEF, Amsterdam 1009, The Netherlands}
\address[kurchatov]{Kurchatov Institute 123182, Moscow, Russia}
\address[compostela]{University of Santiago de Compostela, Santiago de 
Compostela 15706, Spain}
\address[budker]{Institute of Nuclear Physics, Novosibirsk 630090, Russia}
\address[isr]{Institute for Storage Ring Facilities, University of
Aarhus, Aarhus 8000, Denmark}
\address[aarhus]{University of Aarhus, Aarhus 8000, Denmark}
\address[california]{University of California, Irvine 92697,  USA}
\address[capetown]{University of Cape Town, Cape Town 7701, South Africa}
\collab{NA-59 Collaboration}

\newpage

\begin{abstract}

The CERN-NA-59 experiment examined a wide range of electromagnetic
processes for multi-GeV electrons and photons interacting with oriented
single crystals. The various types of crystals and their orientations were
used for producing photon beams and for converting and measuring their
polarisation.

The radiation emitted by  178~GeV unpolarised electrons incident on a 1.5~cm thick Si
crystal oriented in the Coherent Bremsstrahlung (CB) and the String-of-Strings
(SOS) modes was used to obtain multi-GeV linearly polarised photon beams.

A new crystal polarimetry technique was established for measuring the
linear polarisation of the photon beam. The polarimeter is based on the
dependence of the Coherent Pair Production (CPP) cross section in oriented
single crystals on the direction of the photon polarisation with respect
to the crystal plane. Both a 1~mm thick single crystal of Germanium and a
4~mm thick multi-tile set of synthetic Diamond crystals were used as
analyzers of the linear polarisation.

A birefringence phenomenon, the conversion of the linear polarisation of the
photon beam into circular polarisation, was observed. This was achieved by letting the linearly
polarised photon beam pass through a 10~cm thick Silicon single crystal
that acted as a "quarter wave plate"~(QWP) as suggested by N.~Cabibbo et 
al.

\end{abstract}

\end{frontmatter}

\section{Introduction}
\label{intro}

Interest in polarised particle physics has been increasing in recent
years. Significant polarisation effects may appear in the final-state
products providing useful tools for particle studies. 
Photon initiated interactions are attractive probes, however the experimental
study of polarisation observables after photonuclear reactions requires
intense beams of polarised high energy photons. 
One compelling motive to generate these intense,
highly polarised high energy photon beams comes from the need to
investigate the polarised photo-production mechanisms. For example, the
so-called ``spin crisis of the nucleon'' and its connection to the gluon
polarisation has attracted much attention~\cite{compass,ric,bosted}. For
these purposes both linearly and circularly polarised photon beam
lines with high intensity and energy are required. A well known method to
produce circularly polarised photons is from the interaction of
longitudinally polarised electrons with crystalline or amorphous media,
where the emitted photons are circularly polarised due to conservation of
angular momentum~\cite{olsen}. Calculations based on~\cite{nadz,armen}
show that CB and channeling
radiation in crystals by aligned incidence of 
longitudinally polarised electrons
are also circularly polarised. These effects
can be used to enhance the number of
high energy circularly polarised photons. The real difficulties here are
associated with the production of polarised electrons. Currently, the
highest energy available for polarised electrons is only 
45~GeV~\cite{slac}.

The goals of NA59 collaboration were:
\begin{itemize}
\item Production of linearly polarised photon beams by the CB of
unpolarised electrons in aligned single crystals oriented in the Point
Effect~(PE) and Strings-of-Strings~(SOS) modes.
\item Establishing  a new fast
polarimetry technique using the aligned crystal method.
\item Investigating the birefringent properties of aligned crystals.
This involved studying the conversion of linear to circular polarisation
by using an aligned single crystal as a quarter-wave plate for
photons of energy 100~Gev. The feasibility of producing high
energy circularly polarised photon beams starting from an unpolarised
electron beam was investigated.
\end{itemize}

The detailed description of the experimental setup, data analysis and
results obtained are given in references~\cite{na59-cb,na59-l4,na59-sos}.
In this paper we present the main results devoted to the polarisation
measurement.

The NA59 experiment was performed in the North Area of the CERN SPS, where
unpolarised electron beams with energies above 100\,GeV are available. The
various data sets were obtained with an electron beam with 
an energy of 178\,GeV and an angular divergence of
48\,$\mu$rad ($\sigma$) and 33\,$\mu$rad ($\sigma$) in the horizontal and
vertical planes, respectively.

The experiment was performed in two stages. The linear polarisation of the
photon beam was studied in the first stage, see reference~\cite{na59-cb}. In
the second stage~\cite{na59-l4}, the QWP crystal was
introduced between the radiator and analyser crystals to investigate the
transformation of the linear polarisation of the photon beam into circular
polarisation. The linear polarisation measurements are important
because the technique of identifying circular polarisation is related
to a reduction in linear polarisation and the conservation of
polarisation.

We have been working toward testing the conjecture that it is possible to
produce circularly polarised photon beams in proton accelerators using the
extracted unpolarised high energy secondary electron beams with energies  up to
250\,GeV (CERN) and 125\,GeV (FNAL)~\cite{propos}. These unpolarised
electron beams can produce linearly polarised photons via CB radiation in
an aligned single crystal. One can transform the initial linear
polarisation of the photons into circular polarisation by using the
birefringent properties of aligned crystals. This method
was first proposed by Cabibbo and collaborators in the
1960's~\cite{cabibbo1}. The detailed theory of birefringence in aligned
crystals is given in~\cite{maish1,maish2,strakh1}.

The theoretical predictions showing the possibility of transforming the
linear polarisation of a high energy photon beam into circular
polarisation in the 80-110\,GeV energy range are presented
in~\cite{maish1,strakh1,akop}. The calculations of the energy
and the orientation dependence of the indices of refraction were performed
using the quasi-classical operator method and CPP formulae.
In these references, the optimum thickness for a QWP Si
crystal was found to be 10\,cm. The relevant geometrical parameters
involved the photon beam forming an angle of 2.3\,mrad with the $\langle 110 \rangle $ axis
such that the photon momentum is also directly in the $(110)$ plane of the Si single
crystal. In this case the angle between the photon momentum and crystal plane is
$\psi$=0. For this choice of parameters, the fraction of surviving photons
is 17-20$\%$.

The production of polarised high energy photon beams enables the
development of methods for the determination of the polarisation of the
photon beam. Several polarimetry methods for linear and circular polarised
photon beams have been developed and used in experiments in the last
decades~\cite{potyl}. Historically, pair conversion in single crystals was
proposed, and later successfully used as a method to measure linear
polarisation for photons in the 1-6 GeV range~\cite{barbiellini}. It was
predicted theoretically and later verified experimentally that the pair
production~(PP) cross section and the sensitivity to photon polarisation
in the single crystals increases with increasing energy. Therefore, at
sufficiently high photon energies, a new polarisation technique based on
this effect can be constructed. Due to the large analysing power of the
crystal this technique become competitive to other techniques such as pair
production in amorphous media and photonuclear methods.

The photon polarisation is conveniently expressed using the Stoke's
parametrisation with the Landau convention, where the total elliptical
polarisation is decomposed into two independent linear
components~($\eta_1$ and $\eta_3$) and a circular component~($\eta_2$).  
The NA-59 collaboration used a new crystal method for 
measuring the linear polarisation of the photon
beam. The polarisation dependence of the PP cross section and
the birefringent properties of crystals are key elements of the photon
polarisation measurement. The imaginary part of the refraction index
is related to the PP cross section. This cross section is
sensitive to the relative angle between a crystal plane of a specific
symmetry and the plane of linear polarisation of the incident photon. In
essence, the two orthogonal directions where these two planes are either
parallel or perpendicular to each other yield the greatest difference in
the PP cross section. We therefore studied the pairs created in a
second aligned crystal, called the analyser crystal. A method of
choosing pairs with particular kinematics to enhance the analysing 
power was identified. This has been called the
``quasi-symmetrical pair selection method (y-cut)''~\cite{ycut} and it 
has been used for constructing the pair asymmetry 
(between the parallel and perpendicular configurations) in the PP analysis.
As a result of such a cut, although the total number of events decreases, 
the relative statistical error diminishes. 
This is because it is inversely correlated with the measured
asymmetry. The asymmetry analysis technique is given 
in reference~\cite{na59-cb} in detail.

The types and orientations of the single crystals used in the experiment
are summarised in Table 1.

\begin{table}
\caption {Different types of crystals and their orientations used in the 
experiment}
\renewcommand{\arraystretch}{1.}
\begin{center}
\begin{tabular}{|c|c|c|c|c|c|}
\hline
Crystal        &Radiation  &Purpose     &Axes,      &Orientation            &Thickness     \\
Type           &     Type  &            &Planes     &                       &   \\ 
\hline
\hline

Si            &CB      &Radiator    &$\langle 001\rangle $,      &$\theta_0$=5mrad,                       &1.5cm      \\
              &        &            &(110)                    &$\psi_{(110)}$=70$\mu$rad               &      \\
\hline

Si            &CB     &Quarter Wave   &$\langle 110\rangle $,   &$\theta_0$=2.29mrad,                &10cm    \\
              &       &Plate       & (110)                    &$\psi_{(110)}$=0                &    \\
\hline

Ge           &CB      &Analyser    &$\langle 110\rangle $,   &$\theta_0$=3mrad,               &1mm    \\
              &        &            &(110)                    &$\psi_{(110)}$=0              &      \\
\hline

Diamond      &CB      &Analyser    &$\langle 001\rangle $,   &$\theta_0$=6.2mrad,               &4mm    \\
              &        &            & (110)                   &$\psi_{(110)}$=560$\mu$rad               &      \\
\hline

Si            &SOS      &Radiator    &$\langle 001\rangle $,      &$\theta_0$=0.3mrad,                &1.5cm      \\
              &        &            & (110)                   &$\psi_{(110)}$=0                &      \\
\hline

Diamond      &SOS      &Analyser    &$\langle 001\rangle $,   &$\theta_0$=6.2mrad,               &4mm    \\
              &        &            &(110)                    &$\psi_{(110)}$=465$\mu$rad               &      \\
\hline
\end{tabular}
\end{center}
\end{table}

\section{Experimental Setup}
\label{setup}

\begin{figure}[htbp]
    \centerline{\includegraphics[width=1\textwidth]{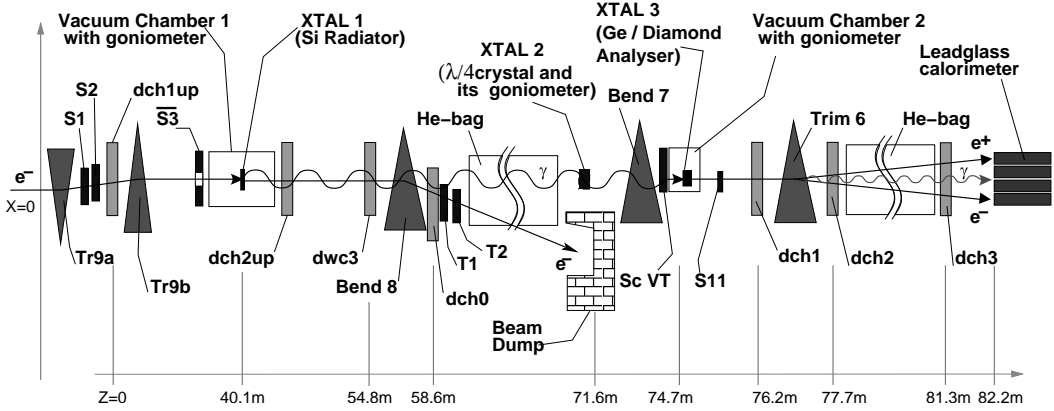}}
    \caption{NA59 experimental setup.}
    \label{F:setup}
\end{figure}

The experimental setup shown in Fig.~\ref{F:setup} was used to investigate
the linear polarization of CB and birefringence in aligned single
crystals~\cite{na59-cb,na59-l4,na59-sos}. This setup is ideally suited for
detailed studies of the photon radiation and pair production processes in
aligned crystals.

The main components of the experimental setup are: three goniometers with
crystals mounted inside vacuum chambers, a pair spectrometer, an electron
tagging system, a segmented leadglass calorimeter, wire chambers, and
plastic scintillators. 

In more detail a 1.5\,cm thick Si crystal can be rotated in the first
goniometer with 2\,$\mu$rad precision and serves as radiator. The second
goniometer needed to control the 10\,cm thick Si crystal, that served as a
QWP, is located after the He-bag. A multi-tile synthetic diamond and
Germanium crystals on the third goniometer can be rotated with
20\,$\mu$rad precision and are used as the analyzer of the linear
polarization of the photon beam.

The photon tagging system consists of a dipole magnet B8, chamber dch0,
and scintillators T1 and T2. Given the geometrical acceptances and the
magnetic field, the system, tags the radiated energy between 10\% and 90\%
of the electron beam energy.

The e$^+$e$^-$ pair spectrometer consists of dipole magnet Trim 6 and of
drift chambers dch1, dch2, and dch3. The drift chambers measure the
horizontal and vertical positions of the passing charged particles with
100\,$\mu$m precision. Together with the magnetic field in the dipole this
gives a momentum resolution of $\sigma_p/p^2=0.0012$ with $p$ in units
of\,GeV/c. The pair spectrometer enables the measurement of the energy of
a high energy photon, $E_\gamma$, in a multi-photon environment.

Drift chambers dch1up, dch2up, and delay wire chamber dwc3 define the
incident and the exit angle of the electron at the radiator. Signals from
the plastic scintillators S1, S2, $\overline{\textrm S3}$, T1, T2, S11 and
veto detector ScVT provide several dedicated triggers~\cite{na59-cb}.

The total radiated energy $E_{tot}$ is measured in a 12-segment array of
leadglass calorimeter with a thickness of 24.6 radiation lengths and a
resolution of $\sigma_E=0.115~\sqrt{E}$ with $E$ in units of\,GeV.  A
central element of this leadglass array is used to map and to align the
crystals with the electron beam.

A detailed description of the NA59 experimental apparatus can be found in 
reference~\cite{na59-cb}.

\section{Production of Linearly Polarised Photons}
\label{production}

CB in oriented single crystals was chosen as a source of the linearly
polarised photon beam. The (CB) method is a well established one for
obtaining linearly polarised photons starting from unpolarised
electrons~\cite{cbdef1,cbdef2}. The relative merits of different single
crystals as CB radiators have been investigated in the
past~\cite{xtalprops}. The silicon crystal stands out as a good choice due
to its availability, ease of growth, and low mosaic spread.

The NA-59 collaboration chose to use a 1.5~cm thick Si crystal as a
radiator to achieve a relatively low photon multiplicity and reasonable
photon emission rate. Two types of crystal orientations, those of the PE and
the SOS orientations, were examined for producing linearly polarised photons.

In the so-called point effect\,(PE) orientation of the crystal, the
direction of the electron beam has a small angle with respect to a chosen
crystallographic plane and a relatively large angle with the
crystallographic axes that are in that plane. For the PE orientation of
the single crystal essentially one reciprocal lattice vector contributes
to the CB cross section~\cite{cbdef1}. The CB radiation from a crystal
aligned in this configuration is more intense than the 
incoherent bremsstrahlung (ICB) radiation in
amorphous media and a high degree of linear polarisation can be achieved.  
The SOS orientation corresponds to the case where the incident
electron momentum lies within a certain crystallographic plane 
making a relatively small angle with one of the crystal axes in that plane.

In the case of the PE orientation of the radiator crystal, the electron beam makes
an angle of 5\,mrad to the $\langle $001$\rangle $ crystallographic axis and about
70\,$\mu$rad from the (110) plane. The resulting photon beam
polarisation spectrum was predicted to yield a maximum polarisation of about
55\% in the vicinity of 70\,GeV and 36\% in the vicinity of 100\,GeV~\cite{na59-cb}.

In the case of the SOS orientation of the radiator crystal, the electron beam
was incident within the $(110)$ plane with an angle of $\theta=0.3$ mrad to
the $\langle 100\rangle $ axis. This hardens the spectrum and 
gives the hard photon peak position (relative to the incident beam energy)
at $x_{max}=0.725$. This corresponds to the photon energy
$E_{\gamma}=125$\,GeV. Under this condition the expected linear
polarisation of photons in the vicinity of 125\,GeV radiation is
negligible~\cite{na59-sos,simon,strakh2}.

\begin{figure}[htbp]
     \includegraphics[width=2.72in]{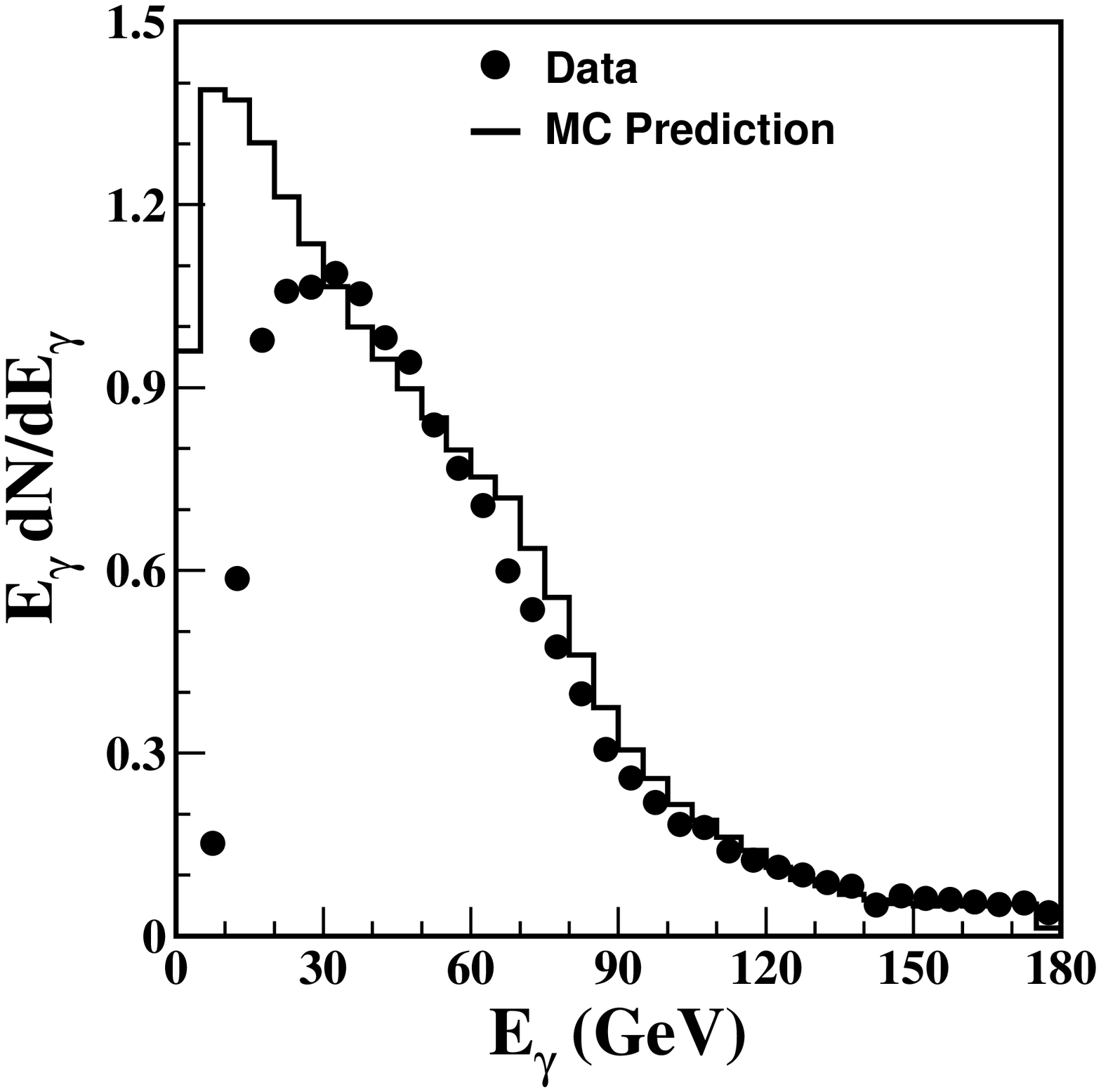}
     \includegraphics[width=2.72in]{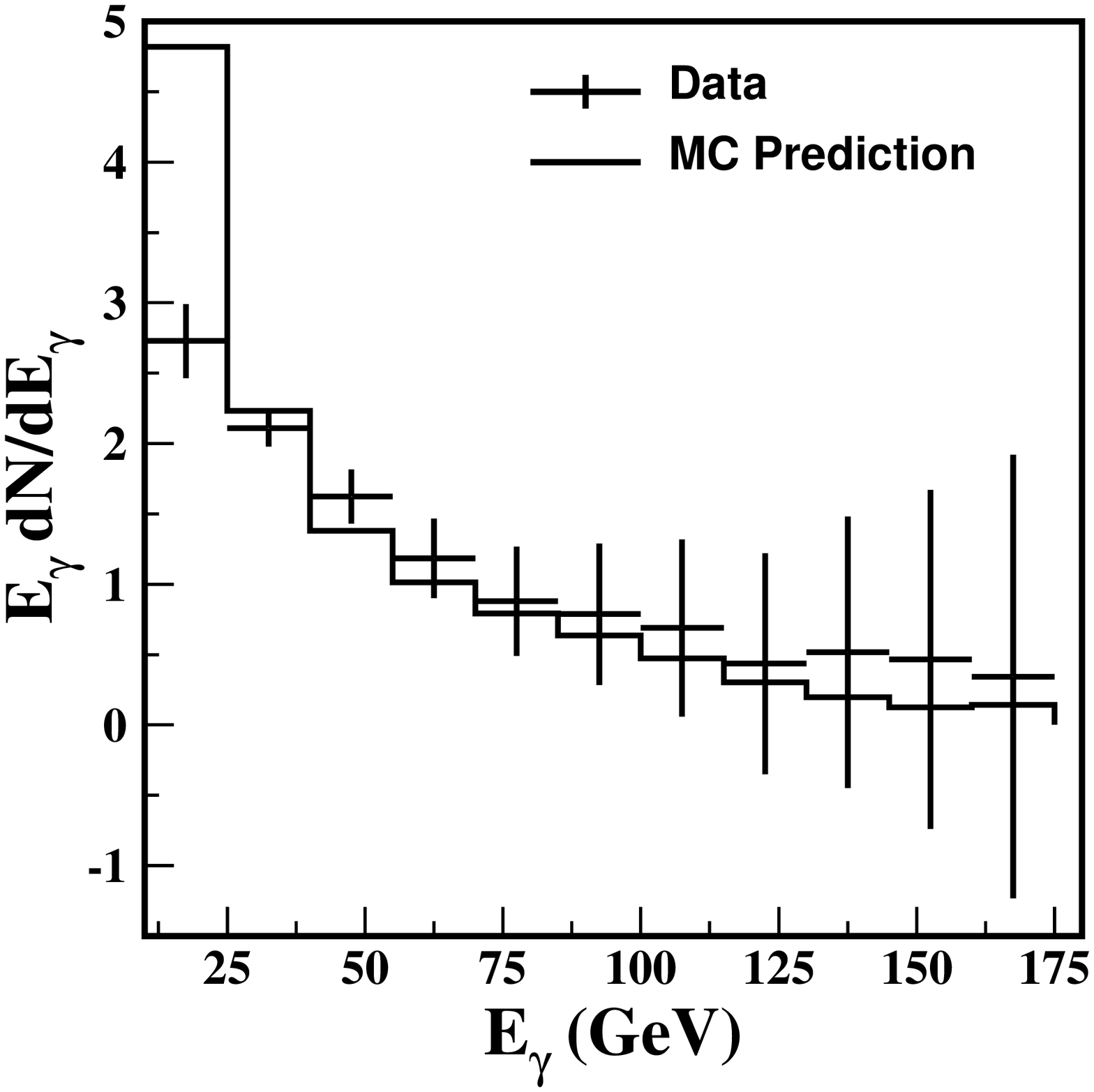}
     \caption{The photon intensity spectra, $E_\gamma dN/dE_\gamma$, 
as a function of the energy $E_\gamma$ of individual photons radiated by 
an electron beam of 178\,GeV in the case of the 1.5\,cm Si crystal aligned 
in the PE~(left) and SOS~(right) modes.
    \label{F:single-g}}
\end{figure}

The single photon intensity spectra are presented in
figure~\ref{F:single-g}. The left figure represents the MC prediction
and the results obtained when the radiator crystal was in the PE
orientation~\cite{na59-cb}.

The right figure represents the MC prediction and the experimental
results obtained when the radiator crystal was in the SOS 
orientation~\cite{na59-sos}. There are several consequences for the
photon spectrum due to the use of a 1.5 cm thick crystal. The photon
multiplicity is above 15~\cite{na59-sos} due to the emission of mainly low
energy photons from planar channeling (PC) for the chosen orientation of
the Si crystal. The most probable radiative energy loss of the 178 GeV
electrons is expected to be 80\%. The beam energy decreases significantly
as the electrons traverse the crystal. The energy of both SOS and PC
radiation also decreases in proportion to the decrease in the electron
energy. Clearly, many electrons may pass through the crystal without
emitting SOS radiation and still lose a large fraction of their energy due
to PC and ICB. Hard photons emitted in the first part of the crystal that
convert in the later part also do not contribute anymore to the high
energy part of the photon spectrum. Consequently, the SOS radiation
spectrum does not show a clearly discernable hard photon peak as would be
the case for a thin radiator~\cite{neweff}. The measured SOS photon
spectrum instead evidences a smoothly decreasing distribution. The low
energy region of the photon spectrum is especially saturated, due to the
abundant production of low energy photons. Above 25\,GeV however, there is
satisfactory agreement with the theoretical MC prediction, which includes
the effects mentioned above.

It follows from the above discussion that CB in crystals aligned in the PE
mode is the more suitable method to increase the intensity of the high
energy part of the gamma spectrum 
(without significant increase of the low energy part)
for tagged photon facilities~\cite{carrigan}.

\section{Results on Measured Asymmetry and Linear Polarisation}
\label{res}

Two types of analyser crystals (Germanium and multi-tile synthetic
diamond), were used in the NA-59 experiment. The selected orientations with
respect to the incident photon beam are given in~Table 1.  
These configurations gave an analysing power peaking
at 90-100~GeV~\cite{na59-cb} and around 125~GeV~\cite{na59-sos} for the PE and
the SOS orientations, respectively. The major advantages of using diamond in
the analyser role are its high pair yield, high analysing power and
radiation hardness. This special  multi-tile synthetic large crystal 
diamond was produced and processed by the  South African
collaborators~\cite{diamond} and mounted and aligned at the ESRF.

\subsection{Linear Polarisation without the Quarter Wave Crystal}

\begin{figure}[h]
      \includegraphics[width=2.72in]{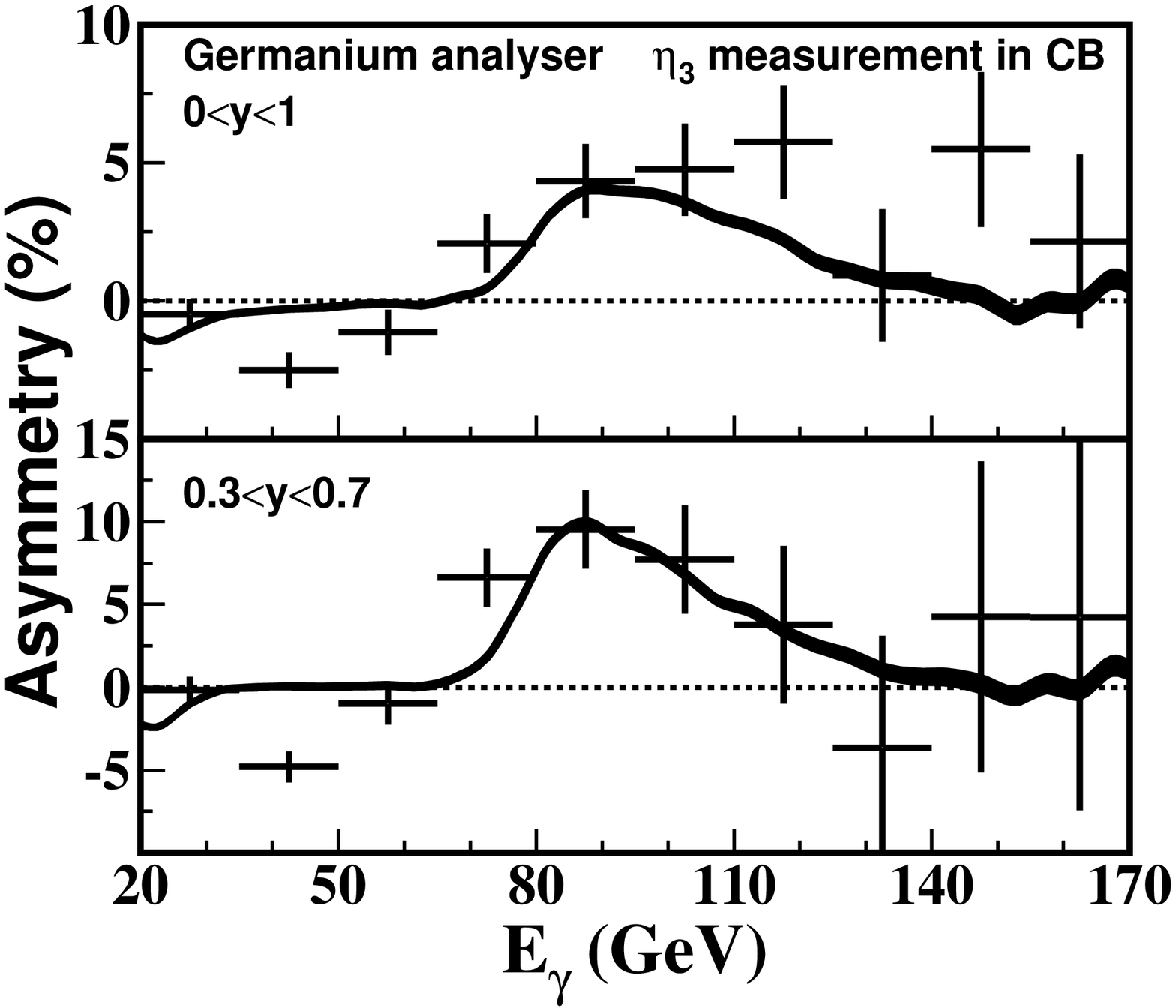}
      \includegraphics[width=2.72in]{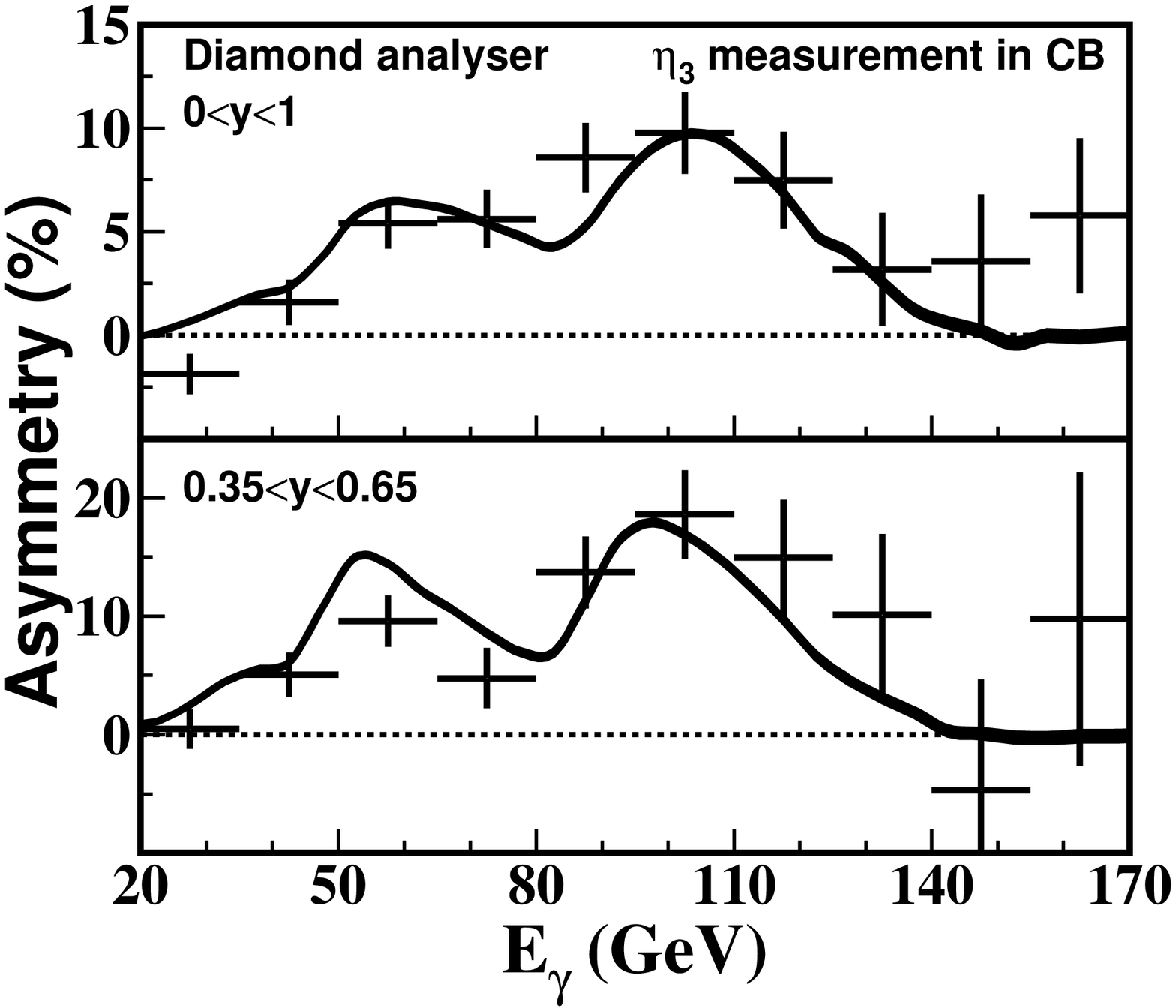}
      \caption{Asymmetry to determine the $\eta_3$ component of the photon 
polarisation with the Ge~(left) and diamond~(right) analysers. 
Measurements without~(top) and with~(bottom) the quasi-symmetrical pair 
selection are presented.
    \label{F:asy-cb}}
\end{figure}

The measured asymmetry in the induced polarisation direction ($\eta_3$) is
presented in figure~\ref{F:asy-cb} with and without the $y$-cut using the
Ge(left) and diamond(right) analyser crystals. The solid line represents
the MC predictions without any broadening effects considered for the
spectrometer. The lower plots represent the increase in the asymmetry due
to quasi-symmetrical pair selection together with the statistical error
associated with this increase. It thus confirms the non statistical source
of the asymmetry increase in the 70-110~GeV range.

Comparing the asymmetry results in figure~\ref{F:asy-cb}, we conclude that
the multi-tile synthetic diamond crystal is a better choice than the Ge
crystal as an analyser, since for the same photon polarisation the former
yields a larger asymmetry and thus enables a more precise measurement. The
diamond analyser also allowed the measurement of the photon polarisation
in the 40-70~GeV range, since it has some, albeit small, analysing power
at these energies.

It is interesting to note that the measured asymmetry in the induced
polarisation direction ($\eta_1$) is consistent with zero~\cite{na59-cb}.

From these results it is easy to determine the Stokes parameter $\eta_3$  
by the formulae (4) from~\cite{na59-cb} using the known analysing power of 
the crystals.

\begin{figure}[htbp]
    \includegraphics[width=2.72in]{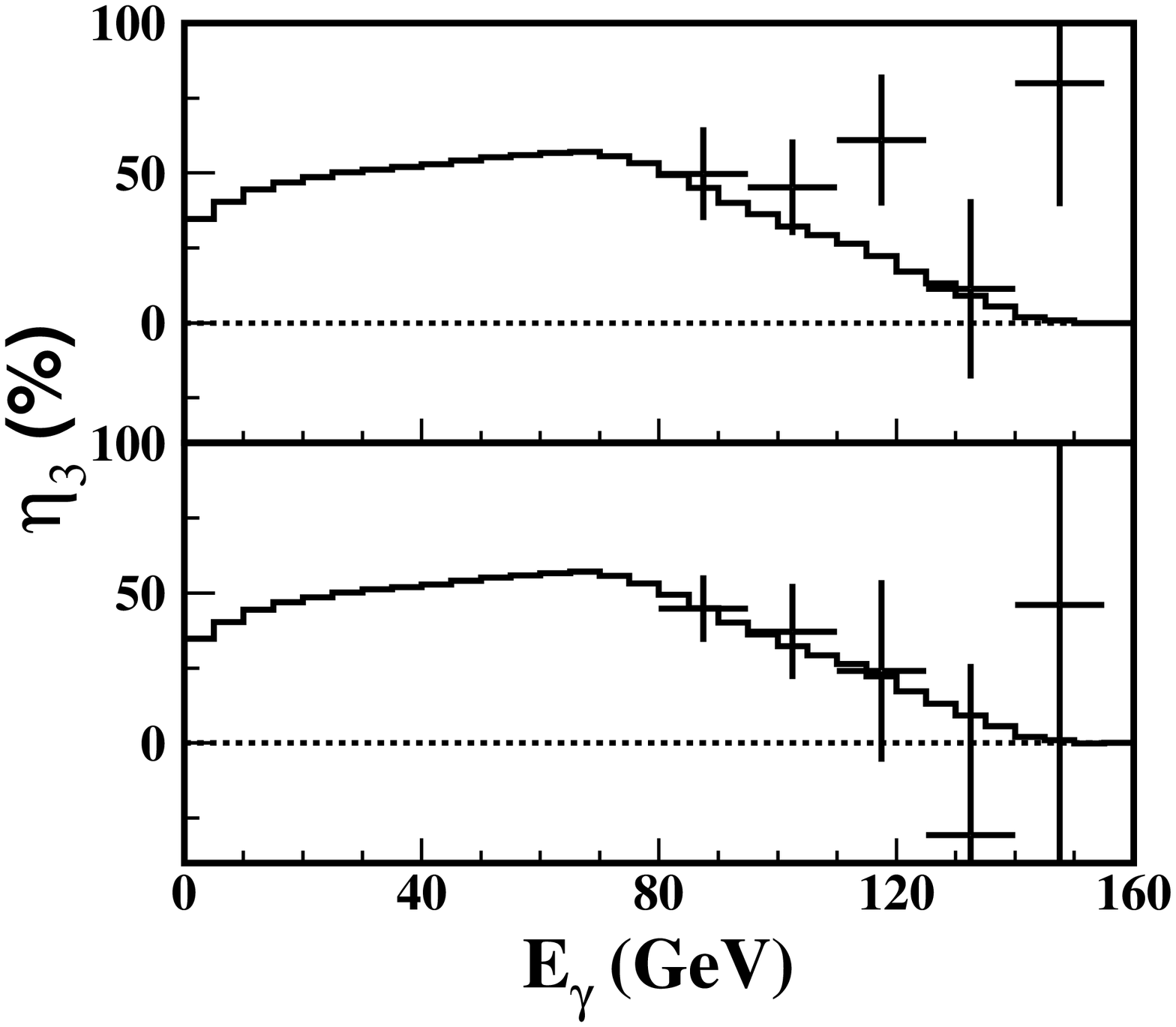}
    \includegraphics[width=2.72in]{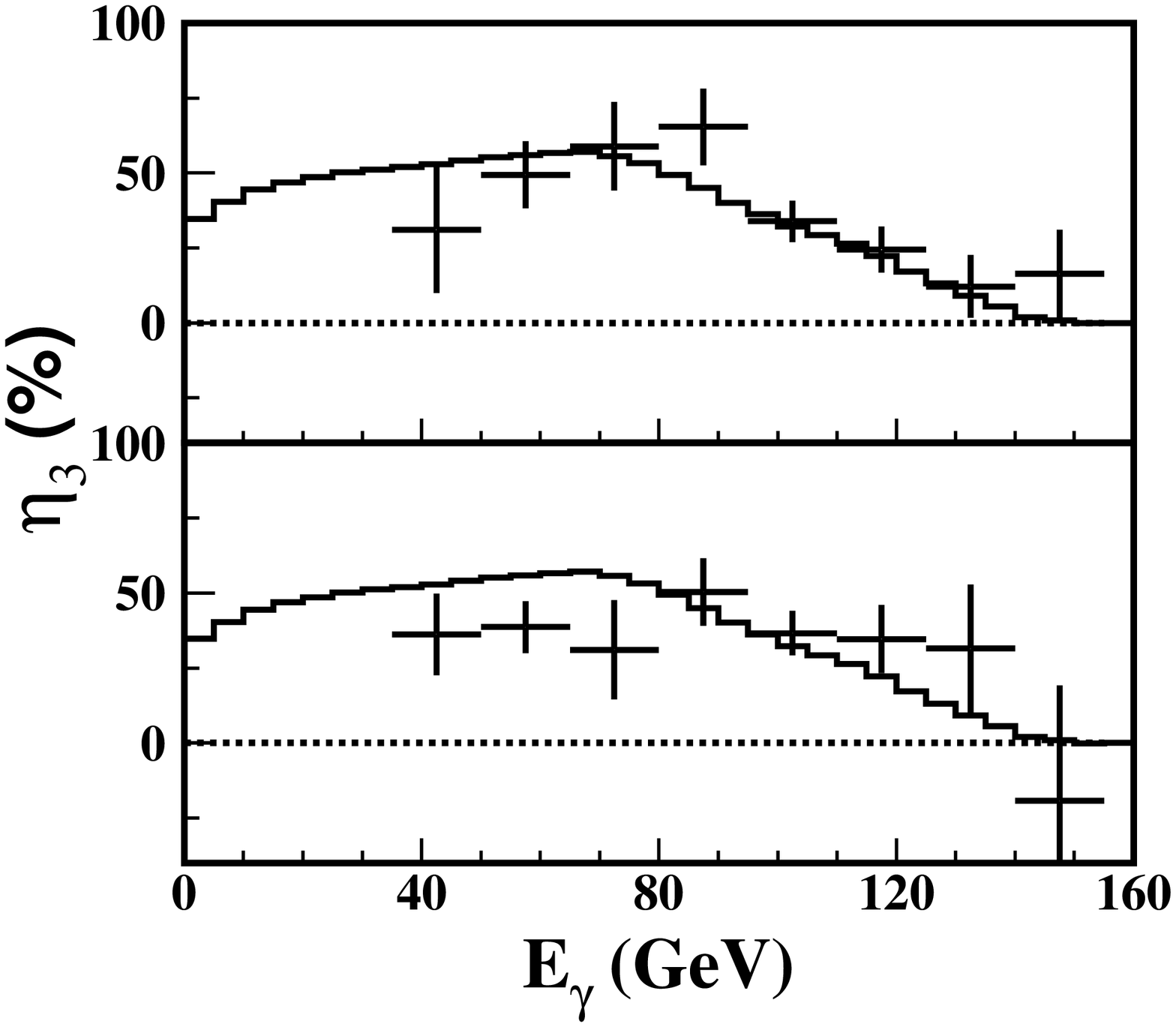}
    \caption{The $\eta_3$ component as a function of photon energy of the 
photon polarisation with the Ge~(left) and diamond~(right) analysers. 
Measurements without~(top) and with~(bottom) the quasi-symmetrical pair 
selection are presented.
    \label{F:pol-cb}}
\end{figure}

The Stokes parameter $\eta_3$ determined by the asymmetry in $e^{-}e^{+}$
PP in both Germanium and diamond analysers is presented in
figure~\ref{F:pol-cb}. The solid lines are the MC
predictions and the crosses are the experimental data. The data for the
$\eta_3$ parameter are
introduced only in the high energy region due to the geometrical
acceptance and smallness of analysing power in the low energy region.

\subsection{Linear Polarisation with the Quarter Wave Crystal}

\begin{figure}[htbp]
      \includegraphics[width=2.72in]{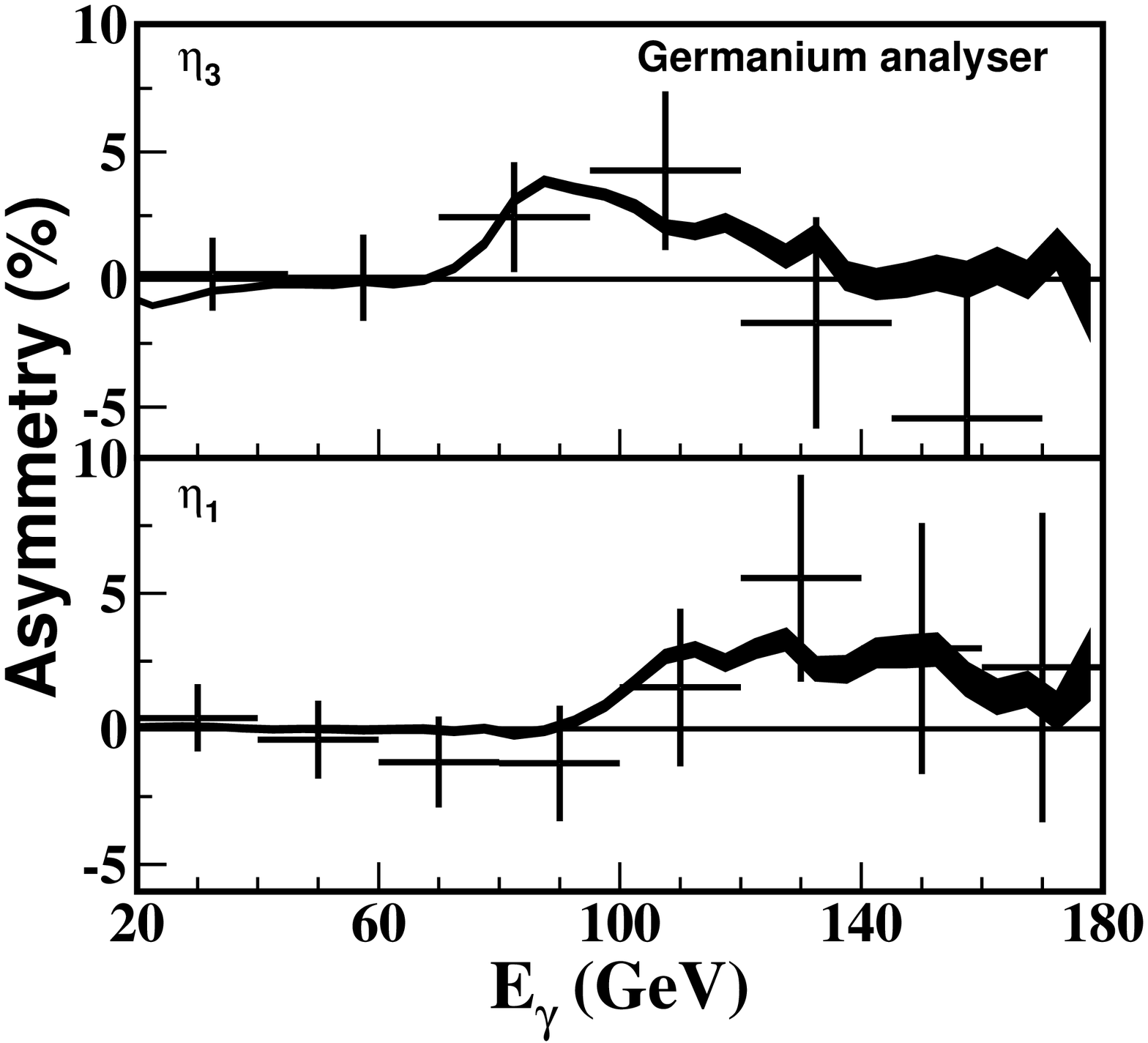}
      \includegraphics[width=2.72in]{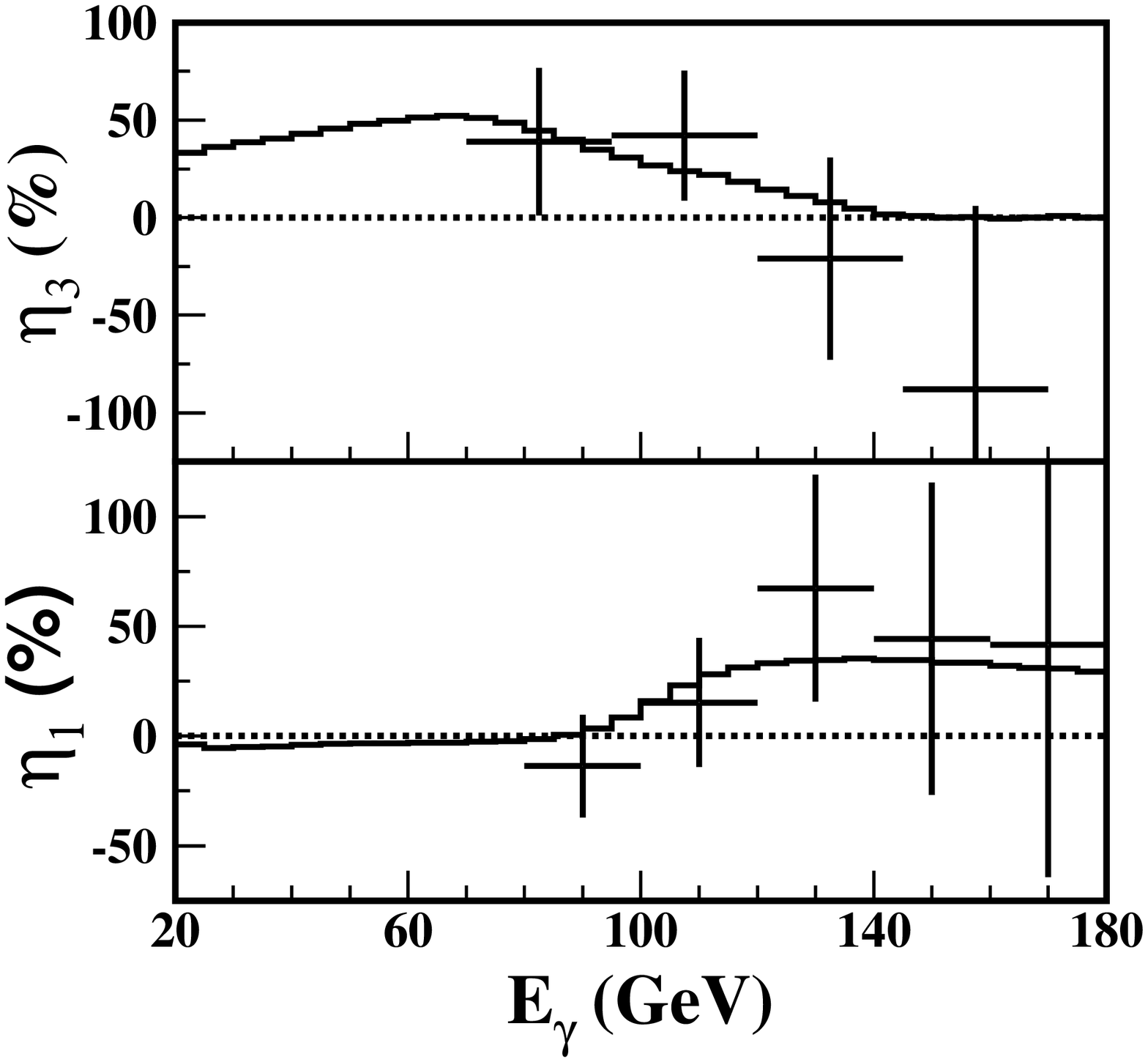}
      \caption{The measured asymmetry~(left) and the Stokes
parameters~(right) as a function of photon energy with the Ge analyser in 
presence of QWP. The top figures are for the  $\eta_3$ component and the 
bottom figures are for the  $\eta_1$ component.
    \label{F:pol-l4}}
\end{figure}

The measured asymmetries~(left) in $e^{-}e^{+}$ PP in the Germanium 
analyser and the Stokes parameters $\eta_1$ and $\eta_3$~(right) are 
presented in figure~\ref{F:pol-l4}. The solid lines are the MC predictions 
and the crosses are the experimental data. The data for the $\eta_1$ and 
the $\eta_3$ parameters are introduced only in the high energy region due 
to the geometrical acceptance and smallness of analysing power in the low 
energy region.

In figure~\ref{F:pol-l4}, we see  an interesting increase of up to a
factor of seven for the $\eta_1$
Stokes parameter in the same energy region. This phenomenon was also
predicted by Cabibbo et al~\cite{cabibbo2}. The unpolarised photon beam
traversing the aligned crystal becomes linearly polarised. This follows
from the fact that the high-energy photons are more strongly affected by the PP
process. The cross section for this depends on the polarisation direction of
the photons with respect to the plane passing through the crystal axis and
the photon momentum (polarisation plane). Thus, the photon beam
penetrating the oriented single crystal feels the anisotropy of the
medium. For the experimental verification of this phenomenon with photon
beams at energies of 9.5\,GeV and 16\,GeV, see~\cite{berger,eisele}. In
the high energy region $>$100\,GeV the difference between the PP cross
sections parallel and perpendicular to the polarisation plane is large.
Since the photon beam can be regarded as a combination of two independent
beams polarised parallel and perpendicular with respect to the
polarisation plane, one of the components will be absorbed to a
greater degree than the other one, and the remaining beam becomes
partially linearly polarised.

\subsection{Linear Polarisation in the SOS orientation}

\begin{figure}[htbp]
      \includegraphics[width=3.2in]{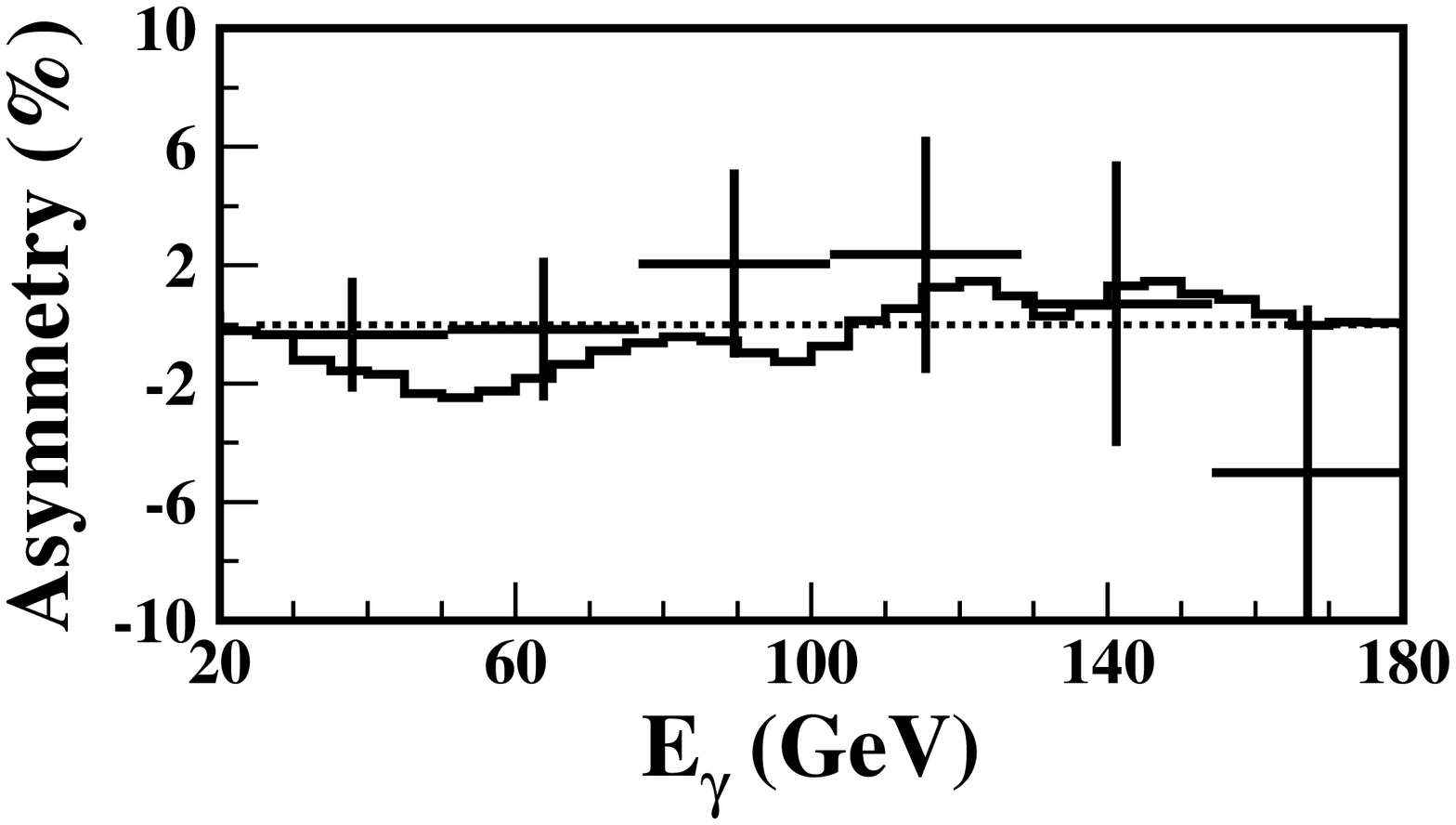}
      \includegraphics[width=2.25in]{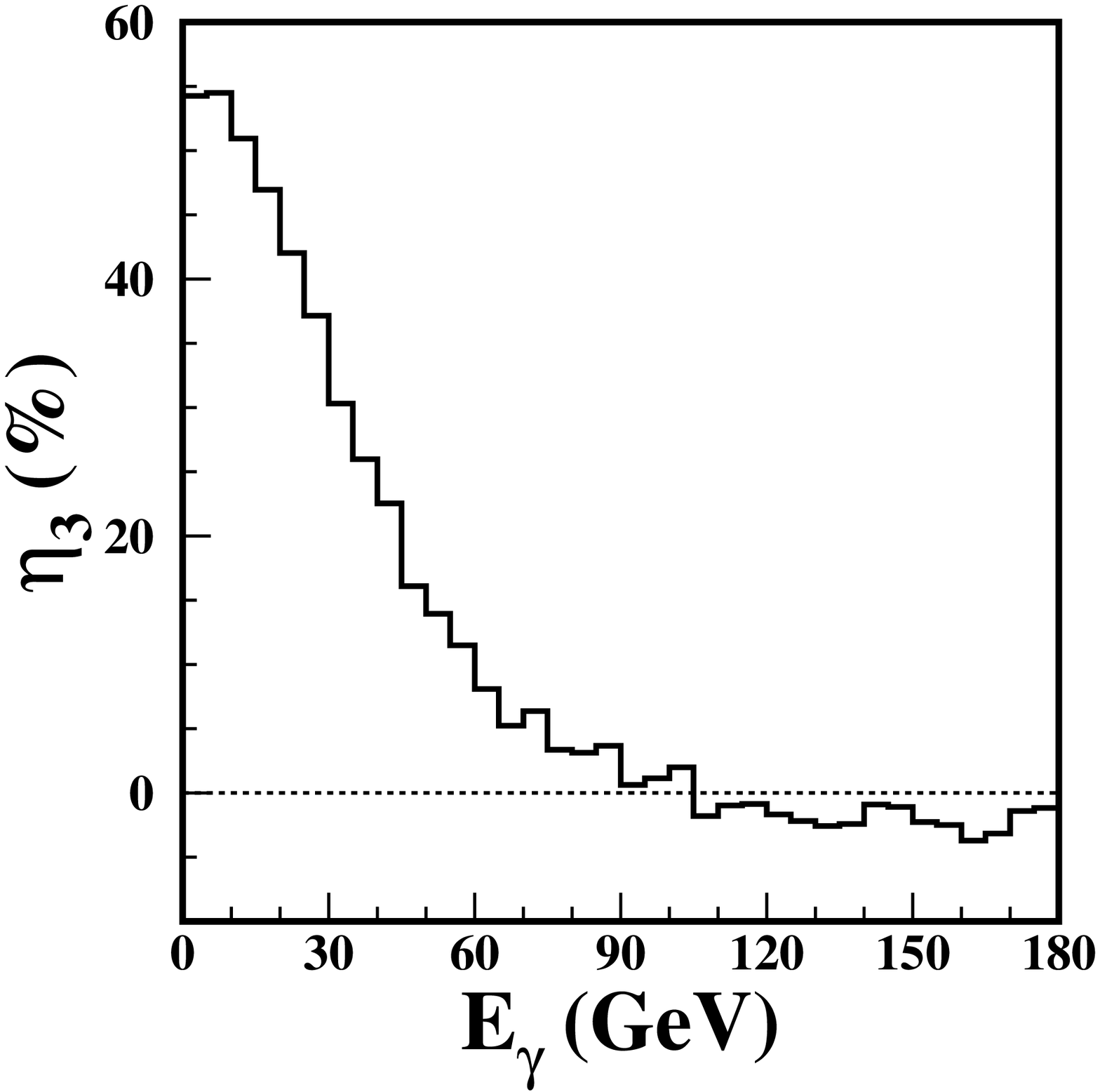}
      \caption{Asymmetry of the e$^+$e$^-$ pair production in the aligned
diamond crystal~(left) and prediction of the $\eta_3$ component~(right)
as a function of the photon energy $E_\gamma$. The black crosses are the 
measurements and the solid line represent the MC prediction.
    \label{F:sos-asy}}
\end{figure}

The measured asymmetry, the predicted asymmetry and the $\eta_3$ 
component of
the linear polarisation are shown in figure~\ref{F:sos-asy}. One can see
that the measured asymmetry is consistent with zero over the whole photon
energy range. The null result is expected to be reliable as the correct
operation of the polarimeter had been confirmed in the same beam-time in
measurements of the polarisation of CB radiation~\cite{na59-cb}. Note,
that the expected asymmetry is small, especially in the high energy range
of 120-140\,GeV, where the analysing power is large~\cite{na59-cb}. This
corresponds to the expected small linear polarisation in the high energy
range, see figure~\ref{F:sos-asy}~(right).

In contrast to the result of a previous experiment~\cite{kirsebom99}, our
results are consistent with calculations that predict negligible
polarisation in the high energy photon peak for the SOS orientation. The
analysing power of the diamond analyser crystal in the previous
experiment's~\cite{kirsebom99} setup peaked in the photon energy range of
20-40\,GeV where a high degree of linear polarisation is expected.  But in
the high energy photon region we expect a small analysing power of about
2-3\%, also following from recent calculations~\cite{simon,strakh2}. The
constant asymmetry measured in a previous experiment \cite{kirsebom99}
over the whole range of total radiated energy may therefore not be due to
the contribution of the high energy photons.

\section{Discussion and Conclusion}
\label{discuss}

The statistical significance of the results was estimated  using
the F-test~\cite{na59-l4,na59-sos}.
These results then show the feasibility of the use of aligned
crystals as linearly polarised high energy photon beam sources. The
predictability of the photon energy and polarisation is a good asset for
designing future beamlines and experiments. These results also establish
the applicability of aligned crystals as polarimeters for an accurate
measurement of the photon polarisation at high energies. The important
aspects are the analyser material selection and utilisation of the
quasi-symmetrical pairs. The use of synthetic diamond as the analyser
crystal is found to be very promising due to its availability, durability
and high analysing power.

Coherence effects in single crystals can be used to transform linear
polarisation of high-energy photons into circular polarisation and vice
versa. Thus, it seems possible to produce circularly polarised photon
beams with energies above 100\,GeV at secondary (unpolarised) electron
beams at high energy proton accelerators. The birefringent effect becomes
more pronounced at higher photon energy, which allows for thinner crystals
with higher transmittance.

We did not perform a direct measurement of the circular polarisation of 
the photon beam. However, realistic theoretical calculations
describe the radiated photon spectrum from the aligned radiator
and the pair production asymmetries in the aligned analyser both with and
without the birefringent Si crystal in the photon beam very well. In view of this
good agreement between the theoretical and predicted linear polarisations for each
case, the predicted birefringent effect seems to be confirmed by
the present measurements. 

Summarising the results, we find the weighted average of the Stokes
parameter before and after the QWP in the energy region 80-110~GeV
$\eta_3$=44$\pm$11\% and $\eta_3$=28$\pm$7\% before and after the QWP
crystal, respectively. From these results one can estimate the $\eta_2$
component of the photon beam polarisation~\cite{na59-l4}, which is
$\eta_2$=21$\pm$11\%. This is consistent with the predicted value of
16\%~\cite{na59-l4}. Given the statistical significance of the data points, 
we find a confidence limit of 73\% for the observation of circular polarisation.

Similar calculations may be done for the Stokes parameter $\eta_1$. If we
make a weighted average for the asymmetry values between 20 and 100 GeV,
where we expect no asymmetry, we obtain a value of 0.19$\pm$0.3\%. Above
100\,GeV we expect a small asymmetry, where we measured (1.4$\pm$0.7)\%.

We also presented new results regarding the features of high energy photon
emission by an electron beam of 178\,GeV penetrating a 1.5\,cm thick
single Si crystal aligned at SOS orientation. This concerns a special case
of coherent bremsstrahlung where the electron interacts with the strong
fields of successive atomic strings in a plane and for which the largest
enhancement of the highest energy photons is expected.  Photons in the
high energy region show less than 20\% linear polarisation at the 90\%
confidence level.

\vspace{1.0cm}

{\bf Acknowledgements}

\vspace{0.5cm}

We dedicate this work to the memory of Friedel Sellschop. We express our
gratitude to CNRS, Grenoble for the crystal alignment and Messers DeBeers
Corporation for providing the high quality synthetic diamonds.  We are
grateful for the help and support of N. Doble, K. Elsener and H. Wahl. We
are grateful to Prof. A.P.~Potylitsin and V.~ Maisheev for fruitful
discussions. It is a pleasure to thank the technical staff of the
participating laboratories and universities for their efforts in the
construction and operation of the experiment.
This research was partially supported by the Illinois Consortium for
Accelerator Research, agreement number~228-1001.


\begin{thebibliography}{00}




\bibitem{compass} G.~Baum et al., Compass Proposal, CERN/SPSLC
96-14,
SPSLC/P297, 1996.
\bibitem{ric} Proposal on Spin Physics Using the RHIC Polarised Collider
(RHIC-Spin Collaboration) 1992; update 1993, (unpublished).
\bibitem{bosted} V.~Ghazikhanian et al., SLAC Proposal 
E-159/160/161, (2000).
\bibitem{olsen} H.~Olsen and L.C.~Maximon, Phys. Rev. 114 (1959) 887.
\bibitem{nadz} I.M.~Nadzhafov, Bull. Acad. Sciencis USSR, Phys. Ser. 14 
(1976) 2248.
\bibitem{armen} A.B.~Apyan, R.O.~Avakian, P.O.~Bosted, S.M.~Darbinian 
and K.A.~Ispirian, Nucl. Instr. and Meth. B145 (1998) 142.
\bibitem{slac} R.~Alley et al., Nucl. Instr. and Meth. A365 (1995) 1.
\bibitem{na59-cb} A.~Apyan et al., NA59 Collaboration, hep-ex/0306028, 
2003.
\bibitem{na59-l4} A.~Apyan et al., NA59 Collaboration, hep-ex/0306041,
2003.
\bibitem{na59-sos} A.~Apyan et al., NA59 Collaboration, hep-ex/0406026, 
2004.
\bibitem{propos} A.~Apyan et al., Proposal to the CERN SPS Committee, 
CERN/SPSC 98-17, SPSC/P308 (1998).
\bibitem{cabibbo1} N.~Cabibbo et al., Phys. Rev. Lett. 9 (1962) 270.
\bibitem{maish1} V.A.~Maisheev, hep-ex/9904029, (1999) 11pp.
\bibitem{maish2} V.A.~Maisheev, V.L.~Mikhalev and A.M.~Frolov, Sov. Phys. 
JETP 74 (1992) 740.
\bibitem{strakh1} V.M.~Strakhovenko, Nucl. Instr. and Meth. B 173 (2001) 
37.
\bibitem{akop} N.Z.~Akopov, A.B.~Apyan and S.M.~Darbinian, hep-ex/0002041 
(2000).
\bibitem{potyl} A.P.~Potylitsin, High Energy Polarised Photon Beams (In 
Russian), Energoatomizdat, Moscow, 1987.
\bibitem{barbiellini} G.~Barbiellini, G.~Bologna, G.~Diambrini and 
G.P.~Murtas, Nuovo Cimento 28 (1963) 435.
\bibitem{ycut} A.B.~Apyan, R.O.~Avakian, S.M.~Darbinian, K.A.~Ispirian, 
S.P.~Taroian, U.~Mikkelsen and E.~Uggerhoj, Nucl. Instr. and Meth. B 173 
(2001) 149.
\bibitem{cbdef1} M.L.~Ter-Mikaelian, High Energy Electromagnetic Processes
in Condensed Media, Wiley Interscience, New-York, 1972.
\bibitem{cbdef2} G.~Diambrini-Palazzi, Rev.\ Mod.\ Phys. 40, (1968) 611.
\bibitem{xtalprops} H.~Bilokon, G.~Bologna, F.~Celani, 
B.~D'Ettorre-Piazzoli, R.~Falcioni, G.~Mannocchi and P.~Picchi, Nucl.\ 
Instrum.\ Meth.\ 204 (1983) 299.
\bibitem{simon} S.M.~Darbinian and N.L. Ter-Isaakyan, Nucl.\ Instr.\ and 
Meth.\ B 187 (2002) 187.
\bibitem{strakh2} V.M.~Strakhovenko, Phys. Rev. A 68, (2003) 042901.
\bibitem{neweff}  R.~Medenwaldt et al., Phys. Lett. B 281, (1992) 153.
\bibitem{carrigan} N.Z.~Akopov, A.B.~Apyan, R.O.~Avakian, R.~Carrigan, 
S.M.~Darbinian, K.A.~Ispirian, Yu.V.~Kononets and S.~Taroyan. Nucl. Instr. 
and Meth. B 115 (1996) 372.
\bibitem{diamond} R.C.~Burns et al., J. Crystal Growth 104 (1990) 257.
\bibitem{cabibbo2}  N.~Cabibbo et al., Nuovo Cimento 27 (1963) 979.
\bibitem{berger} C.~Berger  {\it et al.}, Phys. Rev. Lett. {\bf 25} 
(1970) 
1366.
\bibitem{eisele} R.L.~Eisele, D.~Sherden, R.~Siemann, Charles 
K.~Sinclair, D.J.~Quinn, J.P.~Rutherfoord and M.A. Shupe, Nucl. Instr. and 
Meth. 113 (1973) 489.
\bibitem{kirsebom99}  K.~Kirsebom {\it et al.}, Phys.\ Lett.\ B 459 (1999) 347. 


\end{thebibliography}
\end{document}